\pdfoutput=1
\documentclass[aps,prl,groupedaddress,reprint]{revtex4-1}

\usepackage{amsmath,amssymb}
\DeclareMathOperator\arctanh{arctanh}

\usepackage{hyperref}
\hypersetup{pdfborder={0 0 0},colorlinks=true,urlcolor=blue,citecolor=blue,linkcolor=blue,linktocpage=true}

\begin{document}

\title{Symmetries of post-Galilean expansions}

\author{Joaquim Gomis}
\affiliation{Departament de F\'isica Qu\`antica i Astrof\'isica
and Institut de Ci\`encies del Cosmos (ICCUB), Universitat de Barcelona, Mart\'i Franqu\`es, ES-08028 Barcelona, Spain}

\author{Axel Kleinschmidt}
\altaffiliation{Also at International Solvay Institutes, ULB-Campus Plaine CP231, BE-1050 Brussels, Belgium}
\affiliation{Max-Planck-Institut f\"ur Gravitationsphysik (Albert-Einstein-Institut), Am M\"uhlenberg 1, DE-14476 Potsdam, Germany}

\author{Jakob Palmkvist}
\affiliation{Department of Mathematical Sciences, Chalmers University of Technology and the University of Gothenburg, SE-412 96 G\"oteborg, Sweden}

\author{Patricio Salgado-Rebolledo}
\affiliation{Instituto de F\'isica, Pontificia Universidad Cat\'olica de Valpara\'iso Casilla 4059, Valparaiso, Chile}

\date{\today}

\begin{abstract}
In this letter we study an infinite extension of the Galilei symmetry group in any dimension that can be thought of as a non-relativistic or post-Galilean expansion of the Poincar\'e symmetry. We find an infinite-dimensional vector space on which this generalized Galilei group acts and usual Minkowski space can be modeled by our construction.  We also construct particle and string actions that are invariant under these transformations.
\end{abstract}

\keywords{ICCUB-19-017}

\maketitle

Non-relativistic physics can be obtained from relativistic physics via an expansion in $1/c$ where $c$ is the velocity of light \footnote{In any physical problem, the expansion involves a dimensionless ratio $v/c$ where the velocity $v$ depends on the system under study, e.g. $v$ could be the velocity of a particle or a relative velocity. In theories with extended objects the non-relativistic expansion is not unique \cite{Batlle:2016iel,Barducci:2018wuj}}.
At the level of symmetries, it is well-known that the strict $c\to\infty$ limit of the relativistic Poincar\'e group yields the Galilei group of non-relativistic symmetries via Wigner--In\"on\"u contraction~\cite{Inonu:1953sp}. The non-relativistic gravity theory of Newton is only invariant under the Galilei group and it is not known how to obtain the associated (Newton--Cartan) Lagrangian 
 from the relativistic Einstein--Hilbert action.
 Corrections to the non-relativistic Newtonian theory at higher order in $1/c$
are famously important for the original experimental evidence for general relativity; see, for example, Refs.
\cite{daucourt1964newtonsche,dautcourt1990newtonian} for discussions.
 Besides the expansion in $1/c$ there can be physical situations with multiple parameters that can be small such as the space-time curvature (in units related to the Newton constant). The precise meaning of `post-Newtonian expansion' then depends on which parameters are considered small~\footnote{For isolated self-gravitating systems such as compact binaries these parameters can be related by the virial theorem~\cite{Blanchet:2013haa}. We shall not touch upon post-Minkowskian corrections that are related to expansions in the Newton constant and have attracted attention recently in works that bring QFT perturbation theory methods into the study of gravitational waves.}. Our approach here relies on $1/c$ corrections only and starts from the Galilei symmetry and we therefore refer to our approach as ``post-Galilean". 
Higher-order (parametrized) post-Newtonian corrections to the Keplerian two-body motion are also of central importance in current investigations of binary gravitational wave sources~\cite{Blanchet:1995ez,Buonanno:1998gg,Blanchet:2013haa,Will:2014kxa,Damour:2015isa} and post-Newtonian corrections have also been investigated in the context of cosmology and structure formation, see, for instance, Ref.~\cite{Ehlers:1996wg}. The effective one-body approach~\cite{Maheshwari:1980hn,Buonanno:1998gg} was inspired in part by the developments of the classical mechanical relativistic two-body problem~\cite{Todorov:1976pt,Komar:1978hc,DrozVincent:1978yk,Giachetti:1981gr}.

One common feature of truncated post-Galilean or 
post-Newtonian expansions is that they inherently do not preserve the full relativistic symmetries. As an early example, the quantum-mechanical Breit equation describes corrections of order $v/c$ to the two-body problem for electrons but it is not invariant under Lorentz transformations~\cite{Breit}. 
More recently, due to the possible applications of Newtonian gravities to non-relativistic holography for strongly coupled systems in condensed matter physics,  non-relativistic gravities that are invariant under various classes of non-relativistic symmetry Lie algebras have been investigated~\cite{Andringa:2010it,Hartong:2014pma}. Subsequently, extensions of the Bargmann algebra by a finite number of generators were first presented in Refs.~\cite{Hansen:2018ofj,Ozdemir:2019orp,Bergshoeff:2019ctr} and these describe the next-to-leading orders in the non-relativistic expansion of gravity. The construction can also be generalized to infinite-dimensional algebras~\cite{Khasanov:2011jr,Hansen:2019vqf,Gomis:2019fdh} using different approaches.

In this letter, we exhibit a universal scheme for obtaining 
post-Galilean expansions of non-relativistic systems by means of their symmetry algebras, starting from the Galilei algebra $\mathfrak{G}$. 
Our scheme can be viewed as either stemming from (affine) Kac--Moody algebras~\cite{Gomis:2019fdh} or from a Lie algebra expansion~\cite{Hatsuda:2001pp,deAzcarraga:2002xi,Izaurieta:2006zz}, and it is based on an infinite-dimensional algebra, which we will refer to as $\mathfrak{G}_\infty$. This algebra can be viewed as including the full formal power series in $1/c$ of relativistic systems. The analysis in~\cite{Hansen:2019vqf} is in the context of the large $c$ expansion of general relativity \cite{DePietri:1994je,Dautcourt:1996pm,Tichy:2011te,VandenBleeken:2017rij}, while our approach is purely kinematical and provides an extension of special relativity that describes post-Galilean physics. 
We introduce an infinite-dimensional space on which this symmetry acts and identify subspaces of finite co-dimension that are mapped to one another by the usual action of the Poincar\'e algebra as a specific combination of transformations in $\mathfrak{G}_\infty$. The infinite-dimensional algebra $\mathfrak{G}_\infty$ admits finite-dimensional quotients corresponding to working up to a finite order in $1/c$.
 
The present letter is mainly concerned with outlining the kinematic underpinnings of this symmetry algebra. Truncations of it have been used in a slightly different guise to construct non-relativistic gravity theories~\cite{Hansen:2018ofj,Ozdemir:2019orp,Bergshoeff:2019ctr} that should properly be identified as post-Newtonian gravities. 
In a follow-up paper~\cite{Gomis:2019nih}, we shall study field-theory implementations of the symmetry structure in the framework of three-dimensional Chern--Simons theory
and also the post-Galilean corrections of the non-vibrating non-relativistic string theory \cite{Batlle:2016iel}.

\subsection{\label{sec:infG}
Generalized Minkowski space-time and\\ its infinite-dimensional symmetry}
Our starting point is the $(d+1)$-dimensional
 Poincar\'e algebra $\mathfrak{iso}(d,1)$ written in a non-covariant form
  with generators $\{p_0,j_{ab}, j_{a0}, p_a\}$ and non-trivial commutation relations
\begin{equation}\label{eq:GA}
\begin{aligned}
&[j_{ab}, j_{cd} ] = 4\delta_{[c[b} j_{a]d]}\,,\quad
[j_{ab}, p_c ] = 2 \delta_{c[b} p_{a]}\,,\\
&[j_{a0}, j_{b0} ] = j_{ab}\,, \quad
[p_0, j_{a0} ] = p_a\,,\\
&[j_{a0}, p_{b} ] =- \delta_{ab} p_0\,, \quad
[j_{ab}, j_{c0} ] = 2 \delta_{c[b} j_{a]0}\,.
\end{aligned}
\end{equation}
Roman indices $a=1,\ldots,d$ are vector indices of the spatial rotation algebra $\mathfrak{so}(d)$ that is generated by $j_{ab}$. The other generators will be referred to as time translation ($p_0$), boost ($j_{a0}$), and translation ($p_a$).

 We can construct an associated
infinite-dimensional algebra by applying the method of infinite Lie algebra expansions~\cite{Penafiel:2016ufo,Gomis:2019nih} 
 to the previous algebra. The generators are
 $H^{(m)} = c^{-2m+1} \otimes p_0$, $P_a^{(m)} = c^{-2m} \otimes p_a$, $B_a^{(m)} = c^{-2m-1}\otimes j_{a0}$ and $J_{ab}^{(m)} = c^{2m}\otimes j_{ab}$, which satisfy
 the algebra $\mathfrak{G}_\infty$:
\begin{align}
\label{eq:GG}
[J_{ab}^{(m)}, J_{cd}^{(n)}  ] &= 4\delta_{[c[b} J_{a]d]}^{(m+n)}  \,,\nonumber\\
[ J_{ab}^{(m)}, B_c^{(n)} ] &= 2\delta_{c[b} B_{a]}^{(m+n)}\,, \;
[ J_{ab}^{(m)}, P_c^{(n)} ] = 2\delta_{c[b} P_{a]}^{(m+n)}\,,\nonumber\\
[ H^{(m)}, B_a^{(n)} ] &= P_a^{(m+n)}\,,\;
[ B_a^{(m)} , P_b^{(n)} ] = -\delta_{ab} H^{(m+n+1)}\,,\nonumber\\
[ B_a^{(m)}, B_b^{(n)} ] &= J_{ab}^{(m+n+1)}\,.
\end{align}
This algebra can also be obtained from the positive modes of an affine Kac--Moody~\footnote{This affine Kac--Moody algebra possesses an integer grading different from the superscripts of the generators.} associated to the Galilei algebra~\cite{Gomis:2019fdh}. 

Thinking of the collection of all boost generators $B_a^{(m)}$ and rotation generators $J_{ab}^{(m)}$ for $m\geq 0$ as generating a generalized
Lorentz algebra $\mathfrak{L}_\infty$, we introduce \textit{generalized Minkowski space} as the formal coset space $\exp{\mathfrak{G}_{\infty}} / \exp{\mathfrak{L}_\infty}$.
 We put local coordinates on this infinite-dimensional space by introducing coordinates $t_{(m)}$ and $x_{(m)}^a$ dual to the generalized translation generators $H^{(m)}$ and $P_a^{(m)}$. 

We now consider the action of an infinitesimal transformation of the coordinates on generalized Minkowski space under a general transformation in $\mathfrak{G}_\infty$ of Eq.~(\ref{eq:GG}) with parameters 
$\alpha^{ab}_{(m)}$ (for $J_{ab}^{(m)}$), $\epsilon_{(m)}$ (for $H^{(m)}$), $\epsilon_{(m)}^a$ (for $P_a^{(m)}$) and $v_{(m)}^a$ (for $B_a^{(m)}$). The infinitely many coordinates transform as 
\begin{align}
\label{eq:GGtrm}
\delta x_{(m)}^a &= \epsilon_{(m)}^a +  \sum_{n=0}^m \left(v_{(n)}^a t_{(m-n)}- \delta_{bc}\alpha^{ab}_{(n)} x_{(m-n)}^c\right)\,,\nonumber\\
\delta t_{(m)} &= \epsilon_{(m)} + \sum_{n=0}^{m-1} \delta_{ab} v^a_{(n)} x_{(m-1-n)}^b \,.
\end{align}
Restricting only to the ``zero modes" these transformations take the form of the usual Galilei transformations on $(t_{(0)}, x^a_{(0)})$. However, they differ when including higher modes.

\subsection{Recovering standard Minkowski space}

We now introduce the coordinates
\begin{align}
\label{eq:Memb}
X^a = \sum_{m=0}^\infty c^{-2m} x_{(m)}^a\,,\qquad X^0  =  \sum_{m=0}^\infty c^{-2m+1} t_{(m)}\,,
\end{align}
that are formal power series in $1/c$. Note that the coordinates $t_{(m)}$ have dimension of [L$^{2m}$/T$^{2m-1}$], while the coordinates $x^a_{(m)}$ have dimension of [L$^{2m+1}$/T$^{2m}$]. This ensures that the coordinates $X^0$ and $X^a$ have both dimensions of length and this is compatible with the fact that these coordinates are associated to the Lie algebra generators of Eq. \eqref{eq:GG} that emerge from a particular expansion of the translation algebra in powers of $1/c$.
Expansions of coordinates in $1/c$ have already appeared in the literature in Ref.~\cite{Hansen:2019vqf}. In the following we shall sometimes write $\vec{X}$ for the spatial vector with components $X^a$ and use the dot product to denote scalar products with respect to $\delta_{ab}$.

In order to understand the action of the symmetry algebra~(\ref{eq:GG}) on these coordinates we focus on the boosts~\footnote{We restrict all discussions to boosts for simplicity; the same analysis applies to rotations and translations.} and introduce a similar collective series-expanded parameter
\begin{align}\label{eq:DefTheta}
\theta^a = \sum_{m=0}^\infty c^{-2m-1} v^a_{(m)}\,.
\end{align}
The action of this parameter on the coordinates~\eqref{eq:Memb} becomes from Eq.~\eqref{eq:GGtrm}
\begin{align}
\label{eq:GGB}
\delta  \vec{X} &= 
\vec{\theta} X^0\,,\qquad
 \delta X^0 = \vec{\theta}\cdot \vec{X}\,.
\end{align}
These transformations resemble formally the standard Lorentz boosts except for the fact that $\theta^a$ 
contains an infinity of independent components $v^a_{(m)}$. 
In order to remedy this we let
\begin{align}
\label{eq:stb}
v_{(m)}^a = \frac{1}{2m+1} v^{2m+1}  n^a\,,
\end{align}
with the same unit vector $n^a$ for all $m$ and a single additional parameter $v$. 
By replacing Eq. \eqref{eq:stb} in the definition \eqref{eq:GGB}, we obtain
\begin{equation}\label{eq:infboost}
\begin{aligned}
\delta \vec{X} &= \vec{\theta} X^0 =\sum_{m=0}^\infty \frac{1}{2m+1}  \left(\frac{v}{c}\right)^{2m+1} \vec{n} X^0 \,,\\
\delta X^0 &= \vec{\theta}\cdot\vec{X} =\sum_{m=0}^\infty \frac{1}{2m+1} \left(\frac{v}{c}\right)^{2m+1} \vec{n}\cdot\vec{X}\,
\end{aligned}
\end{equation}
and we recover the usual expression of the rapidity parameter in terms of the boost velocity, i.e., $\theta^a=\theta n^a$ with $\theta=\arctanh(v/c)$ upon using the series expansion of $\arctanh$ \footnote{We also note that at this point we have evaluated the formal power series in $1/c$ and convergence requires $v<c$.}. The linear transformation in $\theta$ in Eq. \eqref{eq:infboost} agrees with the standard infinitesimal Lorentz boost on Minkowski space.

While we have thus arrived at a formal agreement, the definition~(\ref{eq:Memb}) requires additional discussion. For fixed $(X^0,X^a)$ this defines a subspace of generalized Minkowski space of co-dimension $d+1$. 
The transformations~\eqref{eq:GGB} we have just analyzed move between different embedded such subspaces. For general parameters $v_{(m)}^a$ they also move the individual $(t_{(m)}, x_{(m)}^a)$ for $m\geq 0$ around in a non-uniform manner, while Eq.~(\ref{eq:stb}) results in a more uniform transformation of all $(t_{(m)}, x_{(m)}^a)$ as it only uses $d$ independent parameters.
We thus arrive at the conclusion that we should identify standard Minkowski space as the space of subspaces of co-dimension $d+1$ labeled by the coordinates~\eqref{eq:Memb}. We note that identifying Minkowski space by picking local coordinates $(ct_{(0)}, x_{(0)}^a) = (X^0,X^a)$ while setting the ``higher" coordinates to zero is not suitable as the transformations of Eq.~\eqref{eq:GG} either do not preserve this choice or lead to just Galilean boosts instead of relativistic Lorentz boosts.

\subsection{Post-Galilean expansion and finite expansions}

We now compare the structure of the transformations~(\ref{eq:GGtrm}) to those of an expansion in $1/c$ of standard Lorentz boosts. For coordinates ($T,\vec{X})$ on Minkowski space, a finite boost with rapidity parameter $\vec{\theta}=\theta\vec{n}$ and $\theta= \arctanh(v/c)$ acts to order $1/c^2$ by
\begin{align}\label{eq:LBoost1/c2}
T' &= T + \frac1{c} \vec{\theta}\cdot\vec{X} + \frac12  \vec{\theta}^{\,2} T+ \ldots \nonumber\\
&= T+ c^{-2} \vec{v}_{(0)}\cdot\vec{X} + \frac12 c^{-2}  \vec{v}_{(0)}^{\,2} T+O(c^{-3})\,,\\
\vec{X}' &= \vec{X} + c\,\vec{\theta} T  + \frac12 (\vec{\theta}\cdot\vec{X}) \vec{\theta} + \frac{c}6 \vec{\theta}^{\,2}  \vec{\theta} T +\ldots \nonumber\\
&= \vec{X} + \vec{v}_{(0)} T  +c^{-2} \vec{v}_{(1)} T +  \frac{c^{-2}}2(\vec{v}_{(0)}\cdot\vec{X}) \vec{v}_{(0)} \nonumber\\
&\hspace{10mm}+ \frac{c^{-2}}6 \vec{v}_{(0)}^{\,2}  \vec{v}_{(0)} T +O(c^{-3})\,,
\end{align}
where the expansion of 
\begin{align}
\vec\theta=\arctanh(v/c)\vec{n}=\frac{1}{c} \vec{v}_{(0)}  + \frac1{c^3} \vec{v}_{(1)}+\ldots
\end{align}
was introduced. When dropping consistently all terms of higher order, these transformations close.

The step of dropping all higher order terms can be circumvented by formally expanding also the Minkowski coordinates according to
\begin{align}\label{eq:TXexp}
T= t_{(0)} + c^{-2} t_{(1)} +\ldots \,,\quad \vec{X} = \vec{x}_{(0)} + c^{-2} \vec{x}_{(1)} +\ldots
\end{align}
and reading off the coefficients at fixed order in $1/c$. This leads to
\begin{align}
\label{eq:tr1}
t'_{(0)} &= t_{(0)} \,, \qquad \vec{x}_{(0)}' = \vec{x}_{(0)} + \vec{v}_{(0)} t_{(0)}\,,\nonumber\\
t_{(1)}' &= t_{(1)} + \vec{v}_{(0)} \cdot \vec{x}_{(0)} + \frac12 \vec{v}_{(0)}^{\,2} t_{(0)}\,,\\
\vec{x}_{(1)}^{\ \prime} &= \vec{x}_{(1)}\! +\! \vec{v}_{(1)} t_{(0)}\! +\! \vec{v}_{(0)} t_{(1)}\! +\! \frac12\vec{v}_{(0)}^{\,2} \vec{x}_{(0)}\! +\! \frac16 \vec{v}_{(0)}^{\,2} \vec{v}_{(0)} t_{(0)}\nonumber\,.
\end{align}
These finite transformations close without dropping any higher order terms. Moreover, the transformations~\eqref{eq:tr1} coincide exactly with the finite transformations obtained from Eq.~\eqref{eq:GGtrm} when consistently quotienting the algebra $\mathfrak{G}_\infty$ by the ideal spanned by all generators with superscript larger than $N=1$. 
More generally, the algebra~(\ref{eq:GG}) can be truncated consistently by setting to zero all generators with superscript larger than some fixed integer $N\geq0 $. The case $N=0$ corresponds to the Galilei algebra and the first line in Eq. \eqref{eq:tr1} indeed corresponds to the standard Galilean boost on the zero modes while the transformation of $(t_{(1)},\vec{x}_{(1)})$ build up higher polynomials. This pattern persists when increasing $N$ to $N+1$: the lower transformations up to $(t_{(N)}, \vec{x}_{(N)})$ are unchanged and the new $(t_{(N+1)}, \vec{x}_{(N+1)})$ transform with higher degree polynomials. The transformations~(\ref{eq:tr1}) generate a group as the truncation of the algebra (\ref{eq:GG}) to any fixed $N$ is a consistent Lie algebra quotient. We can recover the post-Galilean expansion of the Lorentz boost to any given order $c^{-2N}$ from the transformation of the highest coordinates $(t_{(N)}, \vec{x}_{(N)})$.

The quotient algebra obtained with $N=1$ is exactly the one that has appeared in Ref.~\cite{Hansen:2018ofj} in the context of non-relativistic gravity. Here, we have obtained it from a purely kinematical analysis of post-Galilean expansions.

\subsection{Invariant metric and particle dynamics}
The subspace of the generalized Minkowski space defined for fixed $(X^0,X^a)$ has 
an invariant metric under the transformations (\ref{eq:GGtrm}) given by
\begin{equation}
ds^2=-(d{X^0})^2+ d\vec X ^2\,.
\end{equation}
Using (\ref{eq:Memb}) we have
\begin{equation}\label{eq:expmetric}
ds^2 = \sum_{m,n=0}^\infty  c^{-2(m+n)} \left( -c^2 dt_{(m)}dt_{(n)}  + d\vec{x}_{(m)} \cdot d\vec{x}_{(n)} \right)\,.
\end{equation}
Note that the term of
order $c^0$ is 
\begin{align}
-2dt_{(0)} dt_{(1)} + d\vec{x}_{(0)} \cdot d\vec{x}_{(0)}
\end{align}
and is an invariant metric for the Bargmann algebra, see Eq. \eqref{eq:GGtrm}. We also recognize it as the metric of $(d+2)$-dimensional Minkowski space written in light-cone coordinates. 
This is in agreement with the fact that the Bargmann algebra in $d+1$ dimensions is a subalgebra of the Poincar\'e algebra in $d+2$ dimensions.

From the invariant metric we can construct dynamical systems with 
$\mathfrak{G}_\infty$ symmetry,
 for example,
a particle action for a massive particle by considering 
\begin{equation}
\begin{aligned}\label{eq:partaction}
S_{\text{part}} &
= -m c\int d\tau \sqrt{-\dot{X}^{\mu}\dot{X}_{\mu}} 
=S_{(0)}+S_{(1)}+S_{(2)}+\dots
\end{aligned}
\end{equation}
with 
\begin{align}
S_{(0)}&=  -m \, c^{2}\int d\tau  \, \dot t_{(0)}  \,, \nonumber\\
S_{(1)}&= m \int d\tau  \Bigg[-  \dot t_{(1)} +\frac{ \dot {\vec x}_{(0)}^{\,2}}{2\,\dot t_{(0)}}
\Bigg] \,, \label{actions}\\
S_{(2)}&= \frac{ m}{c^2}\int d\tau  \Bigg[ -\dot t_{(2)} +\frac{ \dot {\vec x}_{(0)}\cdot \dot {\vec x}_{(1)} }{\dot t_{(0)}}
- \frac{\dot t_{(1)} \dot {\vec x}_{(0)}^{\,2}} {2\, \dot t_{(0)}^2} + \frac{ \dot {\vec x}_{(0)}^{\,4}} {8 \,\dot t^3_{(0)} }\Bigg] \,,\nonumber
\end{align}
where $\tau$ is the affine embedding parameter and the dot denotes a $\tau$ derivative. The first action leading to non-trivial dynamics is $S_{(1)}$, which is invariant under the
Bargmann algebra. The variable $t_{(1)}$ can be eliminated, leading to a quasi-invariance under the Galilei algebra 
and to the standard action of the massive non-relativistic particle invariant under worldline diffeomorphisms.

The subsequent actions $S_{(n)}$ with $n>1$ describe non-relativistic particles plus 
post-Galilean corrections; they are individually invariant under Eq. \eqref{eq:GGtrm}.
We will illustrate this by considering the action $S_{(2)}$, which includes the coordinates $(t_{(0)}, \vec{x}_{(0)}, t_{(1)},\vec{x}_{(1)}, t_{(2)})$. Because of the extra coordinate $t_{(2)}$, this action is invariant under a central extension of the symmetry algebra realized by the transformations \eqref{eq:tr1} (this symmetry has been considered in the ($2{+}1$)-dimensional case in Refs. \cite{Ozdemir:2019orp,Bergshoeff:2019ctr}, which is obtained by adding the transformation law
\begin{align}
t^\prime _{(2)}&= t_{(2)}+ \vec v_{(0)} \cdot \vec x_{(1)}+ \vec v_{(1)} \cdot \vec x_{(0)} +   \vec v_{(1)} \cdot  \vec v_{(0)} \, t_{(0)}\nonumber\\
&\hspace{5mm}+ \frac{1}{2} \vec v_{(0)}^{\,2} t_{(1)}+\frac{1}{6} \vec v^{\,2}_{(0)} \vec v_{(0)} \cdot \vec x_{(0)} +\frac{1}{24} \vec v^{\,4}_{(0)}  t_{(0)}\,.
\end{align}
Neglecting the total derivative $\dot{t}_{(2)}$ leads to an action that is only quasi-invariant under the transformations \eqref{eq:tr1}. In fact, its variation gives total derivatives of the form $d(\vec{v}_{(0)} \cdot \vec{x}_{(1)})/d\tau$ and $d(\vec{v}_{(1)} \cdot \vec{x}_{(0)})/d\tau$, which are in correspondence with central extensions 
\begin{equation}
[B_{a}^{(0)},P_{b}^{(1)}]=-\delta_{ab}H^{(2)}\,,\quad
[B_{a}^{(1)},P_{b}^{(0)}]=-\delta_{ab}H^{(2)}\,
\end{equation}
of the algebra truncated at $N=1$.
This is in complete analogy with the analysis of the non-relativistic particle, where the invariance under the Bargmann algebra can be deduced from the quasi-invariance of the action under Galilean transformations, which leads to the central extension $[B_{a}^{(0)},P_{b}^{(0)}]=-\delta_{ab}H^{(1)}$ \footnote{The relation among quasi-invariance
of a Lagrangian and extension of a symmetry algebra was studied in \cite{levy1969group,Marmo:1987rv}.}.

Defining the canonical momenta $\vec p^{\, (m)}=\partial L/\partial{\dot{\vec{x}}_{(m)}}$ and $E^{(m)}=-\partial L/\partial \dot{t}_{(m)}$
for 
$(t_{(0)}, \vec{x}_{(0)}, t_{(1)}, \vec{x}_{(1)}, t_{(2)})$, satisfying the Poisson brackets
\begin{equation}\label{eq:Poissonb}
\{t_{(m)},E^{(n)} \}=-\delta_{m}^{n}\,,\quad \{x^a_{(m)},p_b^{(n)} \}=\delta_{m}^{n}\delta_b^a\,,
\end{equation}
leads to the primary constraints
\begin{equation}
\begin{aligned}
&\vec{p}^{\,(1)}\cdot \vec{p}^{\,(0)} -\frac{1}{2} E^{(1)\,2}-\frac{m}{c^2}\,E^{(0)} = 0
\,,\\
&\frac{1}{2} \vec{p}^{\,(1)\,2} -\frac{m}{c^2}\, E^{(1)} = 0\,,\quad
\frac{m}{c^2}-E^{(2)}=0\,.
\end{aligned}
\end{equation}
These ``generalized mass-shell conditions" define first-class constraints associated with the reparametrization invariance of the action. 
They can be made second-class by introducing a suitable gauge-fixing \cite{henneaux1994quantization}, which in this case can be chosen as 
\begin{equation} \label{eq:gaugefixing}
\frac{1}{c^{2m}}t_{(m)}= t =\tau \quad\text{(for all $m$)}\,,
\end{equation}
including fixing the affine parameter. The symmetries of the corresponding gauge-fixed Lagrangian then become non-linear.
The generalized momenta $E^{(m)}$ are expressed in terms of $\vec{p}^{\,{(m)}}$. Finally by projecting the action on the hyperplane
\begin{equation}\label{eq:hplane}
\frac{1}{c^{2m}}x^a_{(m)}= x^a \quad\text{(for all $m$)}\,,
\end{equation}
 the action $S_{(2)}$ becomes
\begin{equation}
S_{(2)}= \int dt  \Bigg[-mc^2 +\frac{m}{2}  \dot {\vec x}^{\,2} + \frac{m}{8c^2} \dot {\vec x}^{\,4}  \Bigg]  \,.
\end{equation}
Note that the conditions 
\eqref{eq:hplane} break the symmetry (Eq. \eqref{eq:GGtrm}) of the gauge-fixed action. 
The energy and momentum for this action are given  by
\begin{equation}\label{eq:particleEP}
E=mc^2+\frac{m}{2}\dot{\vec{x}}^{\,2}+\frac{3m}{8c^2} \dot{\vec x}^{\,4}\,,\quad
\vec P= m\dot{\vec x} + \frac{m}{2c^2}  \dot{\vec x}^{\,2} \dot{\vec x}\,,
\end{equation}
which correspond to the usual non-relativistic relations of energy and momentum plus their first post-Galilean corrections. 

Note that reparametrization invariance of the action $S_{(2)}$ defined in Eq. \eqref{actions} implies that the corresponding Hamiltonian vanishes. However, the gauge fixing condition \eqref{eq:gaugefixing} breaks this invariance and the gauge-fixed form of $S_{(2)}$ has a non-vanishing Hamiltonian, which reduces to the expression for the energy $E$ in Eq. \eqref{eq:particleEP} after projecting the spatial coordinates as in Eq. \eqref{eq:hplane}. On the other hand, the expansions in $1/c$ given in Eq. \eqref{eq:particleEP} can be recovered from the canonical momenta defined in Eq. \eqref{eq:Poissonb} as $\sum_m c^{2m} E^{(m)}$ and $\sum_m c^{2m} \vec{p}^{\,(m)}$, respectively, after imposing Eqs. \eqref{eq:gaugefixing} and \eqref{eq:hplane}. In the same way, more general actions $S_{(n)}$ incorporate more generalized momenta and first-class constraints, leading to post-Galilean corrections up to order $1/c^{2n-2}$ for a non-relativistic particle \cite{GKPS}.

The procedure outlined above can be used to analyze more general physical systems. For instance, one can use the coordinate expansion \eqref{eq:Memb} in the Nambu--Goto action for a bosonic relativistic string.
In the same way one can consider the Nambu--Goto action
\begin{equation}
\begin{aligned}
S_{\text{string}} &=-\frac{T}{c}\int d\tau d\sigma\left[\left(\dot{X}^{\mu}X_{\mu}^{\prime}\right)^{2}-\dot{X}^{\mu}\dot{X}_{\mu}X^{\prime\nu}X_{\nu}^{\prime}\right]^{1/2}\\
&=  -T\int d\tau d\sigma\Bigg[\sqrt{ \left|\dot{t}_{(0)}\vec{x}_{(0)}^{\prime}-t^\prime_{(0)}\dot{\vec{x}}_{(0)}^{\prime}\right|^{2}}+\dots\Bigg] \,,
\end{aligned}
\end{equation}
where the next-to-leading term, proportional to $1/c^2$, is straightforward to compute but rather cumbersome and therefore omitted here. Unlike the 
case of the relativistic particle, the first term in this expansion is exactly invariant under the Galilei algebra \cite{Batlle:2016iel,Gomis:2016zur},
while the term proportional to $1/c^2$ is also invariant under \eqref{eq:tr1} without the need of adding central extensions.

\subsection{Summary}

In this note, we have outlined a procedure for systematically describing symmetries of post-Galilean expansions through the embedding in an enlarged space. The necessity of working up to a certain order in the $1/c$-expansion is replaced by exactly closing symmetry transformations on an extended space. Invariant non-relativistic particles and strings can be easily considered in this language. The same type of symmetry algebras as considered have appeared in the context of non-relativistic gravity theories~\cite{Hansen:2018ofj,Ozdemir:2019orp,Bergshoeff:2019ctr}. In this context, the results shown in Ref. \cite{Hansen:2019vqf} naturally arise when gauging this infinite-dimensional symmetry or, in other words, from a differentiable manifold whose tangent space is the generalized Minkowski space presented here. This procedure can be implemented in any relativistic system to define actions invariant under truncations of $\mathfrak{G}_\infty$, which incorporate post-Galilean corrections of the corresponding Galilean limit.

\medskip
\begin{acknowledgments}
We acknowledge discussions with Eric Bergshoeff, Roberto Casalbuoni, Jaume Gomis, Niels Obers, Harald Pfeiffer, Diederik Roest and Jan Steinhoff. J.G. acknowledges the hospitality and support of the Van Swinderen Institute where this work was finished. J.G. also has been supported in part by MINECO FPA2016-76005-C2-1-P and Consolider CPAN, and by the Spanish government (MINECO/FEDER) under project MDM-2014-0369 of ICCUB (Unidad de Excelencia Mar\`a de Maeztu). P.S-R. acknowledges DI-VRIEA for financial support through Proyecto Postdoctorado 2019 VRIEA-PUCV.
\end{acknowledgments}

%\bibliographystyle{apsrev4-1}
%\bibliography{postnewt}


\end{document}